\documentclass[twocolumn,showpacs,preprintnumbers,amsmath,amssymb]{revtex4}
 
\usepackage{graphicx}
\usepackage{dcolumn}
\usepackage{bm}
\usepackage{epsfig} 

\begin{document}

\preprint{}

\title{Coherent radio pulses from showers in different media: \\
A unified parameterization} 

\author{Jaime Alvarez-Mu\~niz}
\author{Enrique Marqu\'es}
\author{Ricardo A. V\'azquez}
\author{Enrique Zas}
\affiliation{Departamento de F\'\i sica de Part\'\i culas
and Instituto Galego de F\'\i sica de Altas Enerx\'\i as,\\
Universidade de Santiago de Compostela, E-15782 Santiago
de Compostela, Spain}

\begin{abstract} 
We study the frequency and angular dependences of Cherenkov 
radio pulses originated by the excess of electrons in  
electromagnetic showers in different dense media. 
We develop a simple model to relate the main characteristics
of the electric field spectrum to the properties
of the shower such as longitudinal and lateral development.  
This model allows us to establish the scaling of the 
electric field spectrum with the properties of the medium 
such as density, radiation length, 
Moli\`ere radius, critical energy and refraction index.
We normalize the predictions of the scaling relations to 
the numerical results obtained in our own developed 
GEANT4-based Monte Carlo simulation, and we give 
a unified parameterization of the frequency spectrum and angular
distribution of the electric field in ice, salt, and the lunar
regolith, in terms of the relevant properties of the media. 
Our parameterizations are valid for electromagnetic showers below
the energy at which the Landau-Pomeranchuk-Migdal effect 
starts to be relevant in these media. They also
provide an approximate estimate of radio emission in hadronic showers 
induced by high energy cosmic rays or neutrinos. 
\end{abstract}

\pacs{95.85.Bh, 95.85.Ry, 29.40.-n}

\maketitle

\section{Introduction}

Neutrinos are the most 
suitable candidates to extend astronomical observations to the ultra high energy
(UHE) regime, above $\sim 1$ TeV. Unlike charged particles, they point directly
to the source where they were produced. Unlike neutrons, they are stable. 
Unlike photons, they can penetrate large amounts of matter, and they are 
not attenuated by cosmic backgrounds.    
They are not optimal candidates though, due to  
their small interaction cross section,
and to their small expected fluxes at ultra high energy 
\cite{gaisser95,learned00,halzen02}. 
However these two latter difficulties may be overcome with a large enough 
detector volume of at least $1~{\rm km^3}$ \cite{halzen94}. 

The ``conventional"
technique to achieve a large detector volume exploits the observation of 
Cherenkov light emitted by high energy neutrino-induced muons
and showers in water and ice \cite{andres01,AMANDA,Antares,Baikal,spiering03}. 
An alternative to it is the search for coherent Cherenkov radio pulses 
in the MHz-GHz radio frequency range, produced by neutrino
induced showers in dielectric, transparent, dense media.
When the wavelength of the emitted radiation
is larger than the typical dimensions of the shower the 
emission is coherent. 
The contribution to the electric field from positive and
negative particles would approximately cancel out were it not for
the existence of an excess of electrons over positrons in the 
shower \cite{askaryan62}.
The charge asymmetry arises from photon and electron
scattering on the surrounding medium that sweeps electrons into the
shower, and from positron annihilation in flight that contributes
to the excess charge by terminating positron trajectories.
Due to the coherent nature of the radio emission
the power in radio waves scales as the square of shower energy.

Dense media have the advantage that shower dimensions are 
of the order of meters, and coherence extends to high frequencies, 
typically $\sim$ GHz, where more power is available because the 
coherently radiated electric field is proportional to frequency.  
Besides, large formations of dense, transparent media such as ice 
and salt exist in nature having
attenuation lengths in radio frequencies of a few 
hundred meters \cite{Icebook,SalSA02,vandenBerg_ARENA}. They have the advantage
of being located in low noise environments 
such as the South Pole or deep salt mines. A single antenna in such a 
medium makes a detector
of effective volume of $\sim 0.1~{\rm km^3}$ 
water equivalent \cite{price96,Frichter95}.  
The cost-effectiveness of antennas also adds to the attractiveness of
the radio technique.

Several experiments are searching for neutrinos exploiting the radio 
technique in dense media.
The RICE experiment \cite{RICEdetector} 
is an array of antennas buried in the Antarctic ice cap that has
imposed stringent limits on astrophysical neutrino fluxes 
in the energy range above $\sim 10^{17}$ eV \cite{RICElimits}. 
The Goldstone Lunar Experiment (GLUE) has used two large radiotelescopes 
to search for pulses 
produced by neutrino-induced showers on the Moon's regolith \cite{GLUE00}. 
With no neutrino candidates in 120 hours of observation, GLUE has 
established competitive upper bounds on astrophysical neutrinos of 
energy above $\sim 10^{19}$ eV \cite{GLUElimits}. The 
technique could be of interest to search for ultrahigh energy 
cosmic rays \cite{markov86,alvarez_RADHEP} and is being explored using the 
Westerbork array of radiotelescopes in the Netherlands
\cite{Bacelar_ARENA}. It will
also be exploited by the proposed LUNASKA experiment in Australia
\cite{LUNASKA}.  
A recently approved experiment is ANITA \cite{ANITA_ARENA}, a balloon-borne 
antenna that circles the Antarctic continent scanning for large radio 
pulses. The idea of using 
large salt domes as targets for neutrino interactions has also been 
explored  and there is a proposed
experiment named SalSA \cite{gorham-aspen}, as well as other 
initiatives such as the ZESANA proposal in the Netherlands 
\cite{vandenBerg_ARENA}. 

It is also very important to remark that the
dominant emission mechanism in dense media 
has been experimentally confirmed 
in accelerator measurements
at SLAC, using bunches of bremsstrahlung photons 
as projectiles, and sand and salt as target medium 
\cite{saltzberg_SLAC,gorham_salt}. 
This experiment has tested the theoretical predictions 
originally made by Askar'yan in the 1960's \cite{askaryan62}.

Given the current and future
experimental efforts, the motivation and aim of our work are 
clear: A good understanding and comprehensive characterization of the 
dependence of the radio signal on frequency and observation angle in
different dense media are needed to 
interpret the data that is being collected \cite{RICElimits,GLUElimits},
as well as to evaluate the capabilities and potential of the planned
experiments looking for neutrinos and cosmic rays.

In this paper we show that many of the observables of the electric field 
spectrum emitted by an electromagnetic shower developing in dense
media, scale to a few percent level with several properties  
of the media, such as density ($\rho$), 
radiation length ($X_0$), Moli\`ere radius ($R_M$), 
critical energy ($E_c$), and index of refraction ($n$).  
This scaling allows us to predict the properties of
radio pulses in different media without performing time-consuming 
Monte Carlo simulations. Our work provides further
insight into the close relation between shower development and radio emission,
which may help designing future experiments exploiting 
the radio technique, as well as assessing the potential of those currently
under construction.

This paper is structured as follows. In Section \ref{pulse}
we briefly review radio emission from electromagnetic 
showers. In Section \ref{toy} by means of a simple toy model we relate the
main features of the radiopulse to general characteristics 
of shower development. This allows us to 
obtain the scaling of radiopulse with the relevant parameters of the medium. 
In Section \ref{media} we normalize these scaling relations 
using our own developed GEANT4-based Monte Carlo simulation 
\cite{GEANT4,alvarez03} 
to numerically calculate the electric field spectrum in ice,
salt and the lunar regolith. We also give in this section 
unified parameterizations of our numerical results
to be used in practical applications and establish their
range of applicability.  

\section{\label{pulse}The Radio Pulse Spectrum}

We provide here a short review on Cherenkov radio emission from 
electromagnetic showers for the sake of completitude. Further information
can be found in references \cite{zas91,zas92,alvarez97,buniy02,razzaque01}.

The Fourier time-transform in the Fraunhofer limit of the electric
field radiated by a charge $q$ moving at constant velocity $\vec{v}$
for an infinitesimal time interval $(t_1, t_1 + \delta t)$ starting 
at the position $\vec{r}_1$ is given by \cite{zas92}:

\begin{equation}
\label{EF_1Particle}
  \vec{E} (\omega , \vec{r}) =
  \frac{q \mu_r i \omega}{2 \pi \epsilon_0 c^2} \frac{e^{ikr}}{r}
  e^{i \omega (t_1 - \frac{n}{c}\hat{k} \cdot \vec{r}_1)} \vec{v}_{\bot}
 \delta t \frac{\sin y}{y}
\end{equation}
with $y=\pi\nu\delta t (1-n\beta\cos\theta)$.
Here $\epsilon_0$ ($c$) is the permittivity (speed of light) in vacuum,
$\mu _r$ ($n$) is the relative magnetic permeability (the refraction
index) of the medium, $\omega$ the
angular frequency, $\hat{k}$ is a
unit vector in the direction of observation, and
$\vec{v}_\bot$ is the projection of $\vec{v}$ in the direction perpendicular
to $\hat{k}$. $r$ is the distance from the charge to the observer's
position. It is straightforward to
deduce Eq.~(\ref{EF_1Particle}) from the inhomogeneous
Maxwell's equations in the transverse gauge for a point charge current
\cite{jackson,zas92}, keeping only the $R^{-1}$ radiation term.

Eq.~(\ref{EF_1Particle}) displays 
the most important characteristics of 
the radiation emitted by a charged particle travelling along 
a straight trajectory at a speed larger than the speed of light
in the medium:

\begin{enumerate}

\item{The electric field spectrum rises linearly with frequency, 
and has a $\sin y/y$ 
term, where $y\propto \nu\delta t (1-n\beta\cos\theta)$ that corresponds to  
an angular diffraction pattern peaked at the Cherenkov angle
$\theta_c$ given by  $\cos\theta_c=(n\beta)^{-1}$. The first zeros of 
the diffraction pattern happen at,
\begin{equation}
\label{eq:width}
\Delta\theta\simeq \frac{1}{\nu\delta t \sqrt{n^2\beta^2 - 1}},
\end{equation}
that is the width of the peak around the Cherenkov angle is 
proportional to the inverse of the length of the particle's track, and 
to the inverse of the frequency of observation, as corresponds to the 
diffraction pattern produced by a single slit. 
For observation at the Cherenkov angle
there is no phase factor associated
to the position of the particle along the track and the maximum
in the diffraction pattern is achieved \footnote{Note
that there exists an angle $\theta_c=\arccos(1/n\beta)$
if and only if $v>c/n$ i.e. the particle travels at a speed larger than the
speed of light in the medium.}.
}

\item{The electric field is proportional to the length of the 
particle track perpendicular to the direction of observation, 
i.e. it scales as $v\delta t \sin\theta$.}

\item{The field is $100\%$ polarized in the plane containing
the observation direction and the particle trajectory.}

\end{enumerate} 

In order to calculate the electric field produced by a shower, the
contributions from each particle track, as given by 
Eq.~(\ref{EF_1Particle}), have to be summed taking into account their 
relative phases which stem from their different spacetime positions
and different $\vec{\beta}$. To do this numerically, the tracks can 
be subdivided in small steps so that $\vec{v}$ is approximately constant
in each step.

\begin{figure}[t]
  \begin{center}
    \includegraphics[height=18.5pc]{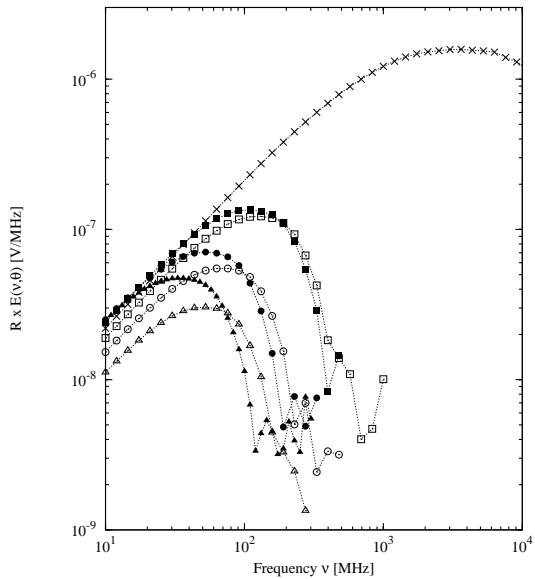}
  \end{center}
  \vspace{-1.5pc}
  \caption{Frequency spectrum of the
electric field in ice observed at several angles from 
the shower axis: 
$\theta=\theta_c\simeq 55.8^\circ$ (crosses);
$\theta_c + 10^{\circ}$ (bold squares); 
$\theta_c - 10^{\circ}$ (empty squares); 
$\theta_c + 20^{\circ}$ (bold circles); 
$\theta_c - 20^{\circ}$ (empty circles); 
$\theta_c + 30^{\circ}$ (bold triangles);
$\theta_c - 30^{\circ}$ (empty triangles).
The results were obtained using our own developed GEANT4-based Monte Carlo
simulation. The lines joining the points are just to guide the eye.
}
\label{freq_spec_angle}
\end{figure}

The contribution to the total electric field from the particle 
tracks in a shower is strongly dependent on 
observation angle and frequency. The problem resembles a classical
problem of interference in which the source emits in a coherent fashion 
as long as the wavelength of observation is insensitive to (i.e. larger than)
the fine details of the emitting region.   
Further information on the characteristics of the electric field
can be found in references
\cite{alvarez97,alvarez98,alvarez99,alvarez00,razzaque01,buniy02}.
In Fig.~\ref{freq_spec_angle}  we show the results of our
own developed GEANT4-based simulation 
of the electric field spectrum emitted by a 10 TeV shower 
in ice \cite{alvarez03}.  The spectrum is plotted
for different observation angles $\theta$ with respect to shower axis. 
The electric field spectrum rises linearly with frequency
up to a turnover point which depends on the
observation angle. 


\section{\label{toy} Toy model of shower development}

In this section we describe a simple toy model of shower 
development that allows
us to relate the main features of the electric field spectrum
to the properties of the shower producing the field.

When a photon or electron of energy $E$ incides on
a thick absorber it initiates an electromagnetic shower.
The number of electrons, positrons and photons in the
shower rapidly grows due to bremsstrahlung
and pair production which are the dominant processes at high
energy. The energy of the primary particle initiating the shower
quickly degrades because it is shared among an increasing
number of secondaries, until their energy reach the
so-called critical energy $E_c$, at which other processes become
important, and electrons are as likely to lose all their energy through
ionization as to radiate hard bremsstrahlung photons. As a result
of the shower development, the number
of particles increases up to a maximum $N_{\rm max}$, and exponentially
decreases after it. Using the Heitler model \cite{heitler}
that assumes that in each interaction the energy is shared
equally between the two secondaries, it is straightforward
to show that $N_{\rm max}=2^{p}=E/E_c$, where $p$ is the number of
interactions before the shower reaches its maximum.
Furthermore, in this model electrons and photons are assumed to
travel a radiation length $L_0=X_0/\rho$ before interacting. As a consequence
the total tracklength $T$ due to the charged particles in a photon
initiated shower can be approximated by \cite{allan71},
\begin{equation}
T \sim (2 + 2^2 + ... + 2^p) L_0
\simeq N_{\rm max}~L_0 = \frac{E}{E_c}\frac{X_0}{\rho}.
\label{track-Heitler}
\end{equation}

\begin{figure}[t]
  \begin{center}
    \includegraphics[height=10.5pc]{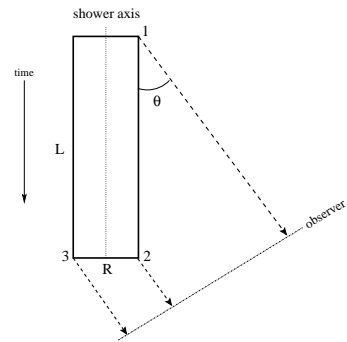}
  \end{center}
  \vspace{-1.pc}
  \caption{Toy model of a shower. $L$ is the
effective length of the shower proportional to the radiation length
$L_0$ and $R$ is the shower width, proportional to the Moli\`ere radius
$R_M$.}
\label{fig:shower_model}
\end{figure}

The particles in the shower also spread in the transverse dimension
-- perpendicular to shower axis -- mainly due to multiple scattering. 
We can view a shower as a thin pancake of 
particles travelling at the speed of light ($\beta=1$) along
a length $L=k_L L_0 = k_L X_0/\rho$ proportional to the radiation
length (Fig.~\ref{fig:shower_model}). The pancake has a lateral width 
$R=k_R R_M/\rho$ proportional to the Moli\`ere
radius $R_M$, the typical
scale of the lateral spread of the shower (Fig.~\ref{fig:shower_model}).
Both $k_L$ and $k_R$ are normalization constants of the model
to be determined in numerical simulations (see below).

This very simple model of shower development, as shown below,
allows us to establish the relation between the frequency,
angular distribution and absolute magnitude of the electric field,
and the parameters of the medium in which it is emitted. 

\subsection{Electric field spectrum: shape}

The overall time duration of the radio pulse for a given observation 
angle from shower axis ($\theta$) is an exceedingly important
parameter. It is related to the turnover frequency at which the spectrum 
departs from a linear behavior with frequency.
There are two relevant time scales in our simple toy 
model, namely, $\delta t_{\rm L}$ 
the time delay between two light rays emitted at the 
two end points in the longitudinal
development of the shower (points 1 and 2 in Fig.~\ref{fig:shower_model}),
and $\delta t_{\rm R}$ the time delay between two rays emitted 
from the two ends of the lateral spread of the shower
(points 2 and 3 in Fig.~\ref{fig:shower_model}). 
The dominance of one or the other
delay -- i.e. the time duration of the pulse -- 
depends on the observation angle. 
It is straightforward to obtain that:
\begin{equation}
\label{eq:delay_long}
\delta t_{\rm L} = \frac{L}{c}~(1-n\cos \theta) =
\frac{n}{c}~\frac{k_L X_0}{\rho}~(\cos \theta_c - \cos \theta),
\end{equation}
where we have used $\cos\theta_c=1/n$, and   
\begin{equation}
\label{eq:delay_lateral}
\delta t_{\rm R} = \frac{R}{c/n} \sin \theta =
\frac{n}{c}~\frac{k_R R_M}{\rho} \sin\theta .
\end{equation}
For observation at the Cherenkov angle all the  
rays emitted  
along $L$ at the same radial distance 
to the shower axis arrive at the same time
at the observer, i.e. $\delta t_{\rm L}$ vanishes
at $\theta=\theta_c$. Only 
the time delay associated to the lateral spread of the shower remains
and the electric field spectrum increases linearly with frequency up
to a turnover frequency given by
\begin{equation}
\label{eq:turnover_lateral}
\nu_{\rm R}(\theta=\theta_c) \simeq \frac{1}{\delta t_{\rm R}} = 
\frac{\rho}{k_R R_M}~\frac{c}{\sqrt{n^2-1}}.
\end{equation}
where we have used $\sin\theta_c=\sqrt{1-\cos^2\theta_c}=\sqrt{n^2-1}/n$.

The corresponding turnover frequency associated to $\delta t_{\rm L}$ 
is,
\begin{equation}
\label{eq:turnover_long}
\nu_{\rm L}(\theta) \simeq \frac{1}{\delta t_{\rm L}} =
\frac{\rho}{k_L X_0}~\frac{c}{\vert 1-n\cos\theta\vert}.
\end{equation}

For observation away from the Cherenkov angle 
both $\delta t_{\rm R}(\theta)$ and $\delta t_{\rm L}(\theta)$ do not vanish.
Eq.~(\ref{eq:delay_long}) illustrates the fact that $\delta t_{\rm L}(\theta)$,
the effective time duration of the longitudinal spread of the shower
as viewed by the observer, grows from zero as the observation angle
departs from $\theta_c$, and equals $\delta t_{\rm R}$  
for observation angles 
$\theta_{+} = \theta_c + \delta\theta_{+}$
and $\theta_{-} = \theta_c - \delta\theta_{-}$ 
given by the condition $\delta t_{\rm L}(\theta_{\pm})=
\delta t_{\rm R}(\theta_{\pm})$.
For small values of $\delta\theta_{\pm}$,  
\begin{equation}
\delta \theta_{\pm} \simeq 
\frac{k_R R_M\sin\theta_c}{k_L X_0\sin\theta_c \mp k_R R_M\cos\theta_c}. 
\label{eq:delta}
\end{equation}
As a result, if $\theta \in (\theta_{-},\theta_{+})$, 
the shower electric field loses its full coherence at   
a turnover frequency $\nu_R$ determined mainly by the {\it lateral} 
spread of the shower and given by an equation similar to 
Eq.~(\ref{eq:turnover_lateral}) but for an arbitrary $\theta$. 
Correspondingly, if $\theta$ falls outside the ($\theta_{-},\theta_{+}$) 
interval, the turnover frequency in the spectrum $\nu_L$ is determined by the 
{\it longitudinal} shower development, and is given by  
Eq.~(\ref{eq:turnover_long}). As a limiting case of Eq.~(\ref{eq:delta}), 
in a shower having $L>>R$, $\delta\theta_{\pm}\sim R/L << 1$   
and the interference from 
different stages in the {\it longitudinal} shower development
dominates the spectrum at essentially all angles, as expected 
in a one dimensional shower where $L$ is the only relevant scale.  

The width of the angular distribution of the electric field  
around the Cherenkov angle is given by an equation similar
to Eq.~(\ref{eq:width}) with an effective $\delta t$ which 
is also proportional to $L$. As a result it is easy to infer that 
it scales as
\begin{equation}
\Delta\theta=\frac{c}{\nu}
\frac{\rho}{k_\Delta X_0}\frac{1}{\sqrt{n^2-1}}.
\label{eq:delta_theta}
\end{equation}
Here $k_\Delta$ is a normalization constant to be determined in Monte Carlo
simulations. Its numerical value is expected to be different 
from $k_L$, despite the fact that both the turnover frequency 
of the spectrum away from the Cherenkov angle, and the angular 
distribution of the pulse are mainly determined by the longitudinal
development of the shower.
 
\subsection{Electric field spectrum: magnitude}

The absolute normalization of the field
at low frequencies -- in the fully coherent region where  
the electric field increases linearly with frequency -- is
known to be determined by the distance travelled by the excess charge
along the perpendicular to the direction of
observation.  Using the expression for the tracklength in 
Eq.~(\ref{track-Heitler}), the absolute 
normalization of the electric field
for an observation angle $\theta$ is expected to scale as
\begin{equation}
r  \vert \vec E\vert 
\sim k_E~\nu~T\sin\theta 
\simeq k_E~\frac{E}{E_c}~\frac{X_0}{\rho}~\nu~\sin\theta,
\label{eq:Efield-norm}
\end{equation}
where $k_E$ is a normalization constant to be determined in 
numerical simulations (see below), and $r$ is the distance
from the source to the observer.
It is important to notice that we are implicitely assuming that 
particles travel parallel to shower axis. As a consequence,
the projection of the tracks in the perpendicular
to the direction of observation is simply accounted for by a 
factor $\sin\theta$ in Eq.~(\ref{eq:Efield-norm}). 

\section{\label{media}Radio Pulse Spectrum in Different Media} 

\subsection{Radio pulse simulations in dense media}

Using our own developed GEANT4-based Monte Carlo code \cite{GEANT4,alvarez03},
we have simulated electromagnetic showers in ice, salt and the lunar
regolith. These media are suitable targets in which neutrino 
interactions can be efficiently detected. It is worth remarking
that we have checked these simulations against the ZHS Monte Carlo
\cite{zas92}, an independent code whose results were
shown in \cite{alvarez03} to agree with those of
GEANT4 to a few percent level in ice. Table \ref{tab:media}
lists the relevant properties of the three media adopted in this work.

\begin{table}[ht]
 \caption{Some relevant properties of the materials considered in this work:
 density $\rho$, 
radiation length $X_0$, Moli\'ere radius $R_M$, critical energy $E_c$,
and index of refraction at radio frequencies $n$. Last column is the 
energy above which the Landau-Pomeranchuk-Migdal effect starts to affect
shower development using the definition in \cite{stanev_LPM}.}
\renewcommand{\arraystretch}{1.5}
\begin{center}
\begin{tabular}{ l | c c c c c c} \hline \hline

Medium  & ~~~$\rho$~~~ & ~~~$X_0$~~~ & ~~~$R_M$~~~ & ~~~$E_c$~~~ & ~~$n$~~ & ~~$E_{\rm LPM}$~~ \\
        & [${\rm g~cm^{-3}}$] & [${\rm g~cm^{-2}}$]  & [${\rm g~cm^{-2}}$] & [MeV] &  & [TeV]   \\ \hline

Ice     & 0.92 & 36.08 & 10.35  & 73.0 & 1.78 & $\sim$ 2400 \\
Salt    & 2.05 & 22.16 & 12.09  & 38.5 & 2.45 & $\sim$ 660 \\
Moon    & 1.80 & 22.59 & 11.70  & 40.0 & 1.73 & $\sim$ 770 \\ \hline \hline
\end{tabular}
\end{center}
\label{tab:media}
\vspace{-1.5pc}
\end{table}
 
\begin{figure}[ht]
  \begin{center}
    \includegraphics[height=14.5pc]{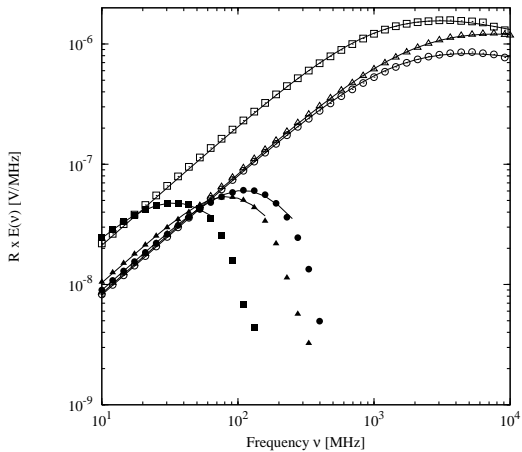}
  \end{center}
  \vspace{-1.5pc}
  \caption{The frequency spectrum of the
electric field emitted by a 10 TeV shower in the direction of  
the Cherenkov angle and at 90 degrees in different dense media.
Empty squares: ice, $\theta=\theta_c$; 
empty circles: salt, $\theta=\theta_c$; 
empty triangles: lunar regolith, $\theta=\theta_c$. 
Filled squares: ice, $\theta=90$ deg. ;
filled circles: salt, $\theta=90$ deg. ;
filled triangles: lunar regolith, $\theta=90$ deg.
The lines are the fits of the Monte Carlo results
to Eqs.~(\ref{eq:param_thetac}) and (\ref{eq:param_theta}), 
using the values of the parameters in Table~\ref{tab:params}.}
\label{EF_media}
\end{figure}

\begin{figure}[ht]
  \begin{center}
    \includegraphics[height=14.5pc]{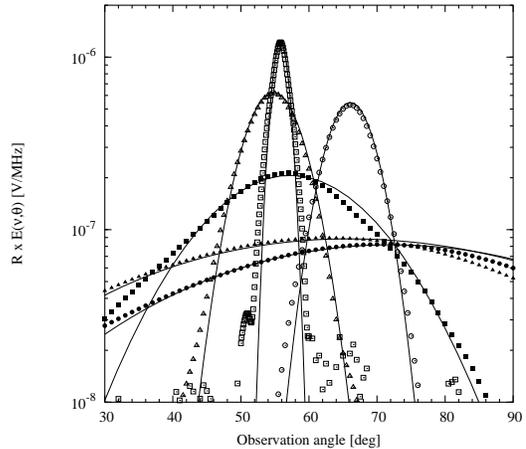}
  \end{center}
  \vspace{-1.5pc}
  \caption{Angular distribution of the electric field 
emitted by a 10 TeV shower in ice, salt and the lunar regolith. The
distributions are shown at $\nu=100$ MHz and $\nu=1$ GHz. 
Empty squares: ice, $\nu=1$ GHz; filled squares: ice, $\nu=100$ MHz.
Empty circles: salt, $\nu=1$ GHz; filled circles: salt, $\nu=100$ MHz.
Empty triangles: lunar regolith, $\nu=1$ GHz; filled triangles: 
lunar regolith, $\nu=100$ MHz.
The lines are the fits of the Monte Carlo results
to Eq.~(\ref{eq:param_ang}) using the values of the 
parameters in Table~\ref{tab:params}.}
\label{AD_media}
\end{figure}

In Fig.~\ref{EF_media} we show the frequency spectrum of the
electric field emitted by a 10 TeV electromagnetic shower observed
at the Cherenkov angle and at $\theta=90$ deg. in ice,
salt and the Moon's regolith, as obtained
in our GEANT4-based simulations. 
In Fig.~\ref{AD_media} we plot the angular distribution
of the electric field at 100 MHz and 1 GHz. It is obvious by
inspection that the shape of the frequency spectrum is fairly universal,
only the normalization and the turnover frequency in the spectrum change 
from medium to medium and scale with its properties, 
as will be quantitatively shown in the next subsection.

\subsection{A unified parameterization}

From Eqs.~(\ref{eq:turnover_lateral}), (\ref{eq:turnover_long}),
(\ref{eq:delta_theta}) and (\ref{eq:Efield-norm}), we have 
obtained a unified parameterization 
of the electric field spectrum and its angular distribution
around the Cherenkov angle in terms of the relevant properties
of the media that control radio emission.
The electric field spectrum at the Cherenkov angle can be 
parameterized by the expression: 
%
\begin{equation}
r \vert{\vec E(\nu)}\vert_{\theta_c}~=~
k_E~\frac{E}{E_c} \frac{X_0}{\rho} ~ \frac{\nu}{\rm 1~MHz} ~\sin\theta_c ~ 
\frac{1}{1+\left({\nu/\nu_R}\right)^\alpha},
\label{eq:param_thetac}
\end{equation}
where $\nu_R$ is given in Eq.~(\ref{eq:turnover_lateral}). 
The parameterization is a good approximation to the electric field
spectrum for frequencies typically below $\nu=$10 GHz where
coherent effects are still very important.

Away from the Cherenkov angle (i.e. for $\theta$ outside the 
interval $(\theta_{-},\theta_{+})$ ) the electric field spectrum is well 
described by:
\begin{equation}
r \vert{\vec E(\nu)}\vert_\theta~=~
k_E~\frac{E}{E_c} \frac{X_0}{\rho} ~ \frac{\nu}{\rm 1~MHz} ~\sin\theta ~ 
\frac{1}{1+\left(\nu/\nu_L\right)^\beta}~,
\label{eq:param_theta}
\end{equation}
%
where $\nu_L$ is given in Eq.~(\ref{eq:turnover_long}).
This expression is generally valid up to frequencies 
$\sim 2 \nu_L$. For larger frequencies, the fine structure
of the shower is relevant (even at the individual particle level),
and clearly our simplified toy model does not account for it.
 
The angular distribution of the pulse around the Cherenkov angle can be 
parameterized by a Gaussian peak modulated by a $\sin\theta$ which 
accounts for the angular behavior in the fully coherent region:
\begin{equation}
r \vert{\vec E(\theta)}\vert = 
r \vert{\vec E(\nu)}\vert_{\theta_c} ~ \frac{\sin\theta}{\sin\theta_c} ~ 
\exp \left[-\left( \frac{\theta-\theta_c}{\Delta\theta} \right)^2\right],
\label{eq:param_ang}
\end{equation}  
with $\Delta\theta$ in radians given in Eq.~(\ref{eq:delta_theta}).

In Eqs.~(\ref{eq:param_thetac}), (\ref{eq:param_theta}) and 
(\ref{eq:param_ang}), $k_E$, $k_L$, $k_\Delta$, $k_R$, 
$\alpha$ and $\beta$ are parameters to be determined 
by fitting the results of Monte Carlo simulations of radiopulse emission
in different media to the parameterizations. 
For this purpose, we have performed fits to the frequency spectrum at the
Cherenkov angle and away from it, as well as to the angular
distribution of the field around the Cherenkov angle, in 
ice, salt and the regolith using Eqs.~(\ref{eq:param_thetac}), 
(\ref{eq:param_theta}) and (\ref{eq:param_ang}),  
and the numerical values of $\rho$, $X_0$, $R_M$, $E_c$ and 
$n$ in Table~\ref{tab:media}.
Table~\ref{tab:params} lists the results of these fits for 
the different media considered in this work. The fits are shown 
in Figs.~\ref{EF_media} and \ref{AD_media}.

\begin{table}[h]
 \caption{Numerical values of $k_E$ 
(in ${\rm V~cm^{-1}~MHz^{-2}}$), $k_L$, $k_\Delta$, 
$k_R$, $\alpha$ and $\beta$ 
after fitting the results of Monte Carlo simulations to Eqs.
(\ref{eq:param_thetac}), (\ref{eq:param_theta}) and (\ref{eq:param_ang}) 
in different media.}
\renewcommand{\arraystretch}{1.5}
\begin{center}
\begin{tabular}{ l | c c c c c c } \hline \hline

Medium & ~$k_E$~ & ~$k_L$~ & ~$k_\Delta$~ &~$k_R$~ & ~$\alpha$~ & ~$\beta$~ \\ \hline

Ice & ~$4.79~10^{-16}$~ & ~23.80~ & ~18.33~ & ~1.42~ & ~1.32~ & ~2.25~ \\ 

Salt & ~$3.20~10^{-16}$~ & ~21.12~ & ~14.95~ & ~1.33~ & ~1.27~ & ~2.60~ \\ 

Moon & ~$3.30~10^{-16}$~ & ~23.74~ & ~17.78~ & ~1.41~ & ~1.23~ & ~2.70~ \\ \hline \hline

\end{tabular}
\end{center}
\label{tab:params}
\vspace{-1.5pc}
\end{table}

As shown in Table~\ref{tab:params}, it is remarkable that the
numerical values of the normalization constants are weakly 
dependent on the medium. This is reflecting that most of the 
dependence on the medium
is already accounted for in the unified parameterizations
through the parameters $\rho$, $X_0$, $R_M$, $E_c$ and $n$.
This is true at $\sim 40\%$ level in $k_E$, and at
$\sim 15\%$ or less in the other normalization constants, which is 
remarkably good given
the simplicity of our toy model of shower development in which the 
parameterizations are based. 

The largest departure from being medium independent 
appears in $k_E$. The reason for this is that we made the simplification 
that particles in the shower follow straight lines parallel to
the shower axis, while it is well known that they   
deviate from the shower axis by an average angle 
which is medium dependent. This dependence is not included in 
Eq.~(\ref{eq:Efield-norm}). As a result, the projection of the tracks
in the perpendicular to the direction of observation is 
not simply accounted for by a factor $\sin\theta$, but some dependence
on the deviation angle should be included in Eq.~(\ref{eq:Efield-norm}). 
The small departures from being medium independent in the other 
normalization constants can be attributed to the fact that an 
electromagnetic shower is better represented by a ``cone" with 
medium-dependent opening angle, 
rather than by a ``cylinder" as depicted in Fig.~\ref{fig:shower_model}. 
This introduces an extra dependence on the medium 
which is not accounted
for in Eqs.(\ref{eq:turnover_lateral}) and (\ref{eq:turnover_long}).

Finally, we are assuming in Eq.~(\ref{track-Heitler}) 
that the electric field scales with the  
tracklength produced by particles with energy above
the critical energy. However the 
electric field is known to scale with the excess charge tracklength, 
which is mainly due to Compton scattering dominating at energies  
below the critical energy. Using the total tracklength
is however not a bad approximation because it scales in roughly the 
same way with medium parameters as the excess tracklength. 
This is confirmed by our 
GEANT4 simulations which predict a ratio of total
to excess tracklengths with almost no dependence on the medium,
of the order of $3.07$, $3.05$ and $2.98$ in ice, 
salt and the lunar regolith, respectively.
 
\subsection{Discussion: Range of applicability of the parameterizations}

Eq.~(\ref{eq:param_thetac}) together with the 
numerical values of the normalization constants
$k_E$, $k_R$ and $\alpha$, describes accurately 
the electric field spectrum at the Cherenkov angle
for frequencies below 10 GHz. This is well above
the frequencies of interest to experiments using ice,
salt and the lunar regolith as targets for neutrino
detection. The parameterization is also approximately valid
at all energies, because the lateral dimension
of the shower -- responsible for the turnover frequency
at the Cherenkov angle -- does not change much with shower
energy as Monte Carlo simulations
of high energy showers in ice have confirmed \cite{alvarez97,alvarez98}.

Away from the Cherenkov angle Eq.~(\ref{eq:param_theta}),   
together with the numerical values of the normalization 
constants $k_E$, $k_L$ and $\beta$, describe accurately
the electric field spectrum up to about twice the frequency
at which the turnover in the spectrum occurs. However, 
contrary to Eq.~(\ref{eq:param_thetac}), 
Eq.~(\ref{eq:param_theta}) is expected to fail at energies
above the energy ($E_{\rm LPM}$) at which the Landau-Pomeranchuk-Migdal
effect \cite{LPM} starts to affect the longitudinal shower development
of the shower. The LPM effect is known to stretch the longitudinal
development of an electromagnetic shower, and to change its effective
longitudinal dimension. The LPM effect also produces a decrease of
the turnover frequency 
in the spectrum as can be easily understood from Eq.~(\ref{eq:turnover_long}) 
and was shown in \cite{alvarez97}. 
In the last column of Table \ref{tab:media} we list 
$E_{\rm LPM}$ in ice, salt and the lunar 
regolith, using the definition of $E_{\rm LPM}$ in \cite{stanev_LPM}. 

The angular distribution of the electric field around the Cherenkov angle
 in Eqs.~(\ref{eq:delta_theta}) and (\ref{eq:param_ang}), is also
determined by the longitudinal development of the shower, and as a 
consequence it is also affected by the LPM effect. The parameterization
in Eqs.~(\ref{eq:delta_theta}) and (\ref{eq:param_ang}) is only 
valid below $E_{\rm LPM}$. It describes 
-- with an accuracy of $\sim 10\%$ -- the
angular distribution of the electric field for frequencies
above $\sim$ 100 MHz, and in the angular range where the pulse falls 
by a factor $\sim$ 2-3 with respect to its value at the peak. 

Our parameterizations are also expected to be a good approximation to
the coherent Cherenkov radio emission in hadronic showers even at energies 
above $E_{\rm LPM}$. Hadronic showers are known to be less affected by 
the LPM effect \cite{alvarez98}, and their efficiency of emission 
of coherent Cherenkov radiation at high energy has been shown to be similar to 
that of electromagnetic showers \cite{McKay_radio}. Besides, at high
energy, a large fraction of the energy of the hadronic shower goes into
its electromagnetic component \cite{alvarez98}. 
Our parameterizations are then expected to be applicable in estimates
of radio emission in most experiments looking for extremely
high energy neutrinos. These experiments are currently concentrating 
most of their efforts on the detection of the hadronic showers produced in  
$\nu_\mu$ charged current interactions, or in the neutral current 
interaction of any neutrino flavor. The observation of the 
electromagnetic shower produced in a $\nu_e$ charged current interaction 
is expected to be very difficult because the LPM effect 
shrinks the angular distribution of the electric field reducing 
dramatically the available solid angle for detection \cite{ARENA_workshop}. 

The important issues in the last two paragraphs above, 
call for simulations of electromagnetic and hadronic showers
with energies above $E_{\rm LPM}$ in different dense media.
We defer this work to a future paper \cite{inprep}.
 

{\bf Acknowledgments:} We thank P. Gorham, S. Razzaque, D. Saltzberg, 
S. Seunarine, D. Seckel and D. Williams
for many discussions on the topic of the paper. This work was supported by
``Xunta de Galicia" (PGIDIT05 PXIC 20604PN), by ``Ministerio de 
Educaci\'on y Ciencia" (FPA 2004-01198) and by FEDER funds.
J.A.-M. is supported by the ``Ram\'on y Cajal'' program.
We thank CESGA, ``Centro de Supercomputaci\'on de Galicia'' for
computer resources. 



\end{document}